\documentclass[10pt, a4paper]{article}
\pdfoutput=1

\usepackage[sc]{mathpazo} 
\usepackage[T1]{fontenc} 
\usepackage{microtype} 

\usepackage[hmarginratio=1:1,top=32mm,columnsep=20pt, left=15mm]{geometry} 
\usepackage{multicol} 
\usepackage[hang, small,labelfont=bf,up,textfont=it,up]{caption} 
\usepackage{booktabs} 
\usepackage{float} 
\usepackage{bookmark}

\usepackage{lettrine} 
\usepackage{paralist} 

\usepackage{abstract} 

\usepackage{titlesec} 
\renewcommand\thesection{\Roman{section}} 
\renewcommand\thesubsection{\Roman{subsection}} 
\titleformat{\section}[block]{\large\scshape\centering}{\thesection.}{1em}{} 
\titleformat{\subsection}[block]{\large}{\thesubsection.}{1em}{} 

\usepackage{amsmath} 
\usepackage{graphicx} 
\usepackage{array}
\usepackage{multirow}
\usepackage{csquotes}
\usepackage{tabularx}
\usepackage[usenames, dvipsnames]{color}
\usepackage{fancyhdr} 
\hypersetup{
  pdfauthor={Cl\'{e}mentine Cottineau et al},
  pdftitle={Paradoxical Interpretations of Urban Scaling Laws},
  pdfsubject={Paradoxical Interpretations of Urban Scaling Laws},
  urlcolor=blue,
}

\title{\vspace{-15mm}\fontsize{24pt}{10pt}\selectfont\textbf{Paradoxical Interpretations of Urban Scaling Laws}} 

\author{
\large
\textsc{Cl\'{e}mentine Cottineau$^{1}$}\thanks{corresponding author, \href{mailto:c.cottineau@ucl.ac.uk}{c.cottineau@ucl.ac.uk}}, \textsc{Erez Hatna$^{2}$},
\textsc{Elsa Arcaute$^{1}$}, \textsc{Michael Batty$^{1}$}\\[2mm] 
\normalsize $^1$ Centre for Advanced Spatial Analysis, University College London, UK \\ 
\normalsize $^2$ Center for Advanced Modeling in the Social, Behavioral and Health Sciences, Johns Hopkins University, USA \\
\normalsize  
\vspace{-5mm}
}
\date{}

\begin{document}

\maketitle 

\thispagestyle{fancy} 


\begin{abstract}

Scaling laws are powerful summaries of the variations of urban attributes with city size. However, the validity of their universal meaning for cities is hampered by the observation that different scaling regimes can be encountered for the same territory, time and attribute, depending on the criteria used to delineate cities. The aim of this paper is to present new insights concerning this variation, coupled with a sensitivity analysis of urban scaling in France, for several socio-economic and infrastructural attributes from data collected exhaustively at the local level. The sensitivity analysis considers different aggregations of local units for which data are given by the Population Census. We produce a large variety of definitions of cities (approximatively 5000) by aggregating local Census units corresponding to the systematic combination of three definitional criteria: density, commuting flows and population cutoffs. We then measure the magnitude of scaling estimations and their sensitivity to city definitions for several urban indicators, showing for example that simple population cutoffs impact dramatically on the results obtained for a given system and attribute. Variations are interpreted with respect to the meaning of the attributes (socio-economic descriptors as well as infrastructure) and the urban definitions used (understood as the combination of the three criteria). Because of the Modifiable Areal Unit Problem (MAUP) and of the heterogeneous morphologies and social landscapes in the cities' internal space, scaling estimations are subject to large variations, distorting many of the conclusions on which generative models are based. We conclude that examining scaling variations might be an opportunity to understand better the inner composition of cities with regard to their size, i.e. to link the scales of the city-system with the system of cities.\\

\end{abstract}


\begin{multicols}{2} 

\section{Introduction}

\lettrine[nindent=0em,lines=3]{U} rban scaling laws are powerful summaries of the variations of urban attributes with city size. Indeed, when considering the variation of an absolute urban quantity $Y$ against total population $P$ in a city $i$, there is almost always a covariation between the two \cite{shalizi2001}, frequently in the mathematical form of a power law

\begin{equation*}
\label{eqscaling}
  Y_i  = a * P_i ^\beta 
  \end{equation*}
 where {\it a} represents a time dependent normalisation constant, and {$\beta$} the scaling exponent under enquiry.\\

Superlinear relationships (i.e. : $\beta > 1$) indicate positive returns to scale, or a relative concentration in the largest cities; whereas sublinearity ($\beta < 1$)  is associated with economies of scale, or a relative concentration in the smallest cities. Linear scaling ($\beta \approx 1$)  means that the quantity per capita is constant across city size. Scaling exponents $\beta$ estimated from empirical data have been interpreted as static or evolutionary properties, respectively by Bettencourt et al \cite{bettencourt2007, bettencourt2012} and Pumain et al \cite{pumain2006}. However, they are subject to variations with respect to urban delineation \cite{arcaute2015}, which questions the validity of a universal interpretation. \\

For example, despite the existence of theoretical models to predict the value of urban scaling from local interactions \cite{bettencourt2012, lobo2013, ortman2014}, an easy way to argue against the universality of scaling exponent values is to look at their variation with city definition. For instance, in France, there are two definitions of cities defined by the statistical office INSEE (cf. table \ref{table:2} and figure \ref{fig:corrdef} in supplement): 

\textbullet { Urban Units or {\it Unit\'{e}s Urbaines} (UU), which correspond to the aggregation of local units ({\it communes}) sharing a continuous built-up area of less than 200m between buildings, and }\\
\textbullet { Metropolitan areas or {\it Aires Urbaines} (AU), defined as the aggregation of a central Urban Unit and all the {\it communes} with more than 40\% of active commuters to the centre.}\\

Comparing scaling results from those two official definitions, we find not only marginal discrepancies between expected values and estimated exponents, but evidence of different scaling regimes when we compare morphological and functional city delineations (cf. table \ref{table:1}), with similar goodness of fits (i.e. quite low for manufacturing jobs and relatively high for the other attributes). In one case, say employment in the manufacturing sector, the number of jobs grows more than proportionally with the population of density-defined Urban Units, whereas the number of such jobs per capita decreases with the size of functionally-defined Metropolitan Areas. \\

\begin{table*}[t]
\begin{center}
\caption{Scaling exponents for two city definitions in France.}
\label{table:1}
\begin{tabular}{|c|c|c|c|c|c|}
\toprule
Urban attribute	& City Definition & $\beta$ & CI ($95\%$) & $R^2$ & N \\
    \hline
    \multirow{2}{*}{Manufacturing}&UU & 1.175 & [1.13;1.22] & 0.543 & 2226\\
    &AU & 0.849 & [0.81;0.89] & 0.691 & 771\\
    \hline
      \multirow{2}{*}{Vacant Dwellings}&UU & 1.051 & [1.03;1.07] & 0.797 & 2233\\
    &AU & 0.902 & [0.88;0.92] & 0.928 & 771\\
    \hline
      \multirow{2}{*}{Basic Services}&UU & 1.086 & [1.07;1.10] & 0.892 & 2233\\
    &AU & 0.956 & [0.94;0.97] & 0.965 & 771\\
    \hline
      \multirow{2}{*}{Education}&UU & 1.215 & [1.19;1.24] & 0.778 & 2230\\
    &AU & 0.981 & [0.96;1.00] & 0.922 & 771\\
\bottomrule
\end{tabular}
\end{center}
Source of the data: French Census, 2011. UU : density-based Urban Units. AU : functionally defined Urban Areas. $N$: Number of cities in the regression.
\end{table*} 

A similarly inconsistent (or paradoxical) result was obtained for $CO_2$ emissions in US cities. In \cite[p.768]{louf2014_epb}, Louf and Barthelemy asked: \textquote{{\it Faced with these two opposite results, what should one conclude? Our point is that, in the absence of a convincing model that accounts for these differences and how they arise, nothing. [...] Conclusions cannot safely be drawn from data analysis alone}}. The paradox obtained from the comparison of city definitions can indeed question the very motivation for using urban scaling and its empirical analysis. However, even though there seems to be no point in trying to fit absolute scaling parameters, the variations in scaling estimation are of theoretical interest because of what they say about the relation between intra-urban spaces (micro-scale), city definitions (meso-scale) and urban scaling (macro-scale).  \\

Indeed, we consider the variations in scaling estimated between dense cities definitions and metropolitan areas not as a failure of a robustness test, but more as the expression of the different nature of urban spaces implied by the two definitions: the former describes the population within a dense environment of social interactions and infrastructure elements; the latter refers to a much larger functional space of economic interactions. Both can be called cities but they are not equivalent. For example, if one was interested in modelling the development of road infrastructures or industry locations, one would consider different strategies in the central and suburban parts of the city, because of differentiated opportunities to locate certain types of buildings (industry for example), because of housing rent gradients or because of the urban atmosphere. Therefore, where the boundary is set to observe cities with respect to scaling is of crucial importance. The boundary concept also applies to the minimum population required to call a population aggregate urban and there might be differences of nature (and quality) between small towns and large metropolises with respect to certain indicators. \\

An additional motivation to explore the multiple city definitions comes from the fact that official definitions rely on the choice of unique thresholds (e.g. distance between buildings, the percentage of commuters or a minimum population). Those have proven useful to describe urbanisation over time, but their precise value contains a share of arbitrariness that we want to evaluate in order to strengthen or question conclusions based on these definitions. Finally, varying definitional criteria will eventually produce a picture of scaling estimates between the two official definitions for France and this will help us understand better the discrepancies observed empirically, as well as to compare studies performed on a large number of cities with analyses which analyse the upper part of the urban hierarchy only. \\

This paper is devoted to the analysis of why and when we observe a transition from one scaling regime to another through generating a whole range of city definitions; in other words, by rearranging in multiple ways the local Census units that compose the different definitions of cities (section \ref{sec:p1}). We analyse the variation of urban scaling with regard to the criteria used to define such cities, and argue that variations are not random (section \ref{sec:p21}). Instead, they can inform our knowledge of cities and the different parts they are composed of. We suggest a way to describe how discrepancies appear as the observed cities vary in their spatial and population extent. We finally propose potential explanations to help understand better the inner composition and morphology of cities at the two geographical levels (section \ref{sec:p22}). Section \ref{sec:p3} concludes by focusing our interest on using urban scaling along with complementary explanations (regional, path-dependent, etc.) of the genesis of city systems. 


\section{A Multiple Representation of the French system of cities}
\label{sec:p1}

\textquote{{\it The extent of the city is important in a number of respects, not least in relation to the question of city size, an issue of considerable significance in urban and regional analysis}}\cite[p.381]{parr2007}. Since our analysis relates to the variation of attributes with city size, knowing the sensitivity of this variation with respect to the spatial criteria for urban delineation is an important aspect. Beyond sensitivity, the different parts of the city (centre and periphery for example) are not expected to behave similarly vis-a-vis economic and infrastructural patterns. Indeed, they are composed of very different populations, built environments and lifestyles, hence representing an internal diversity of physical and social landscapes within the city \cite{guilluy2013}. The inclusion of some or all of these urban components might strongly affect our estimation of scaling laws. This argument is usually left out of the scope of predictive scaling theories where cities are considered as homogeneous objects (for example in \cite{bettencourt2012}). We present three criteria used to delineate cities in France (section \ref{sec:p11}), and the resulting urban clusters generated from combining them in a systematic way (section \ref{sec:p12}), before comparing results with classical definitions (section \ref{sec:p13}).

\subsection{Systematic criteria for defining cities} 
\label{sec:p11}

 Our "continuous'' delineation of cities adapted from Arcaute et al \cite{arcaute2015} follows a similar strategy to the one used in the official identification of cities in France by the statistical office (INSEE), that is : identifying a centre based on a density criteria to aggregate local units, delineating the periphery functionally associated with that centre, based on travel-to-work patterns of each local unit, and eventually applying a minimum population cutoff. The dense centres are called Urban Units (UU), and group communes sharing the same built-up area with a maximum distance of 200m between buildings. The metropolitan areas (AU), correspond to the aggregation of an UU concentrating more than 1500 jobs and with the communes sending more than $40\%$ of their commuters to the UU. In order to perform a sensitivity analysis of urban scaling, we consider a variety of values for each criterion:\\

\textbullet { {\bf Urban density} is associated with historical and morphological centres (or "core cities"), and characterises the part of a city in which the concentration of interactions and economic activity is maximised \cite{parr2007}. Densities can be expressed in relation to the number of buildings or persons per unit of surface. We choose the latter option here, and let the minimum number of residents per hectare define dense cities varying from 1 (loose centres) to 20 (very dense city cores). Population densities of official UUs are distributed within this range, with the median half of cities in the interval $[0.9 ; 3.2]$ with the mean equal to 5.} \\

\textbullet { Over the last two or three decades though, density alone has failed to reflect the spatial extent of urban labour and housing markets. Therefore, it is common to take into account functional indicators of urban activities taking place beyond dense city cores \cite{parr2007, guerois2002}. To define such urban aggregates, researchers usually consider {\bf commuting patterns} to the centres. The criterion for aggregation can refer to the share of income earned in the city core or the share of active residents commuting to the city centre. We choose the latter, and explore the variation of urban extent when this proportion varies from 0 (where a single commuter is sufficient to attach a commune to the urban centre) to $100\%$ (where the functional city basically corresponds to the dense city).} \\

\textbullet { Finally, the {\bf population minimum} plays a major role in the definition of cities, and affects many urban measures (most notably Zipf's exponent, cf. \cite{malecki1980, guerin1995, cristelli2012}). This parameter reveals the conceptual trade-off between acknowledging that cities appear above a critical mass of population and using the larger scope of city sizes in the study of systems of cities (when we use scaling measures). We present results obtained with population cutoffs from 0 (all clusters are considered) to 50,000 inhabitants.} \\

 \subsection{Resulting "urban" clusters}
 \label{sec:p12}
 
The clustering algorithms, developed and described in \cite{arcaute2015}, first group contiguous local units that together exceed the population density criterion $C$. A commune is then linked to the core that attracts its largest percentage of commuters if this percentage is above the flow cutoff $F$. By combining multiple values of $C$, $F$ and a population minimum $P$, this process leads us to consider thousands of representations of the system of cities in France.\\

\begin{figure*}
 \begin{center}
  \caption{A wide range of representation of systems of cities in France ($P=0$)}.
      \label{fig:defs}
\includegraphics[width=\textwidth]{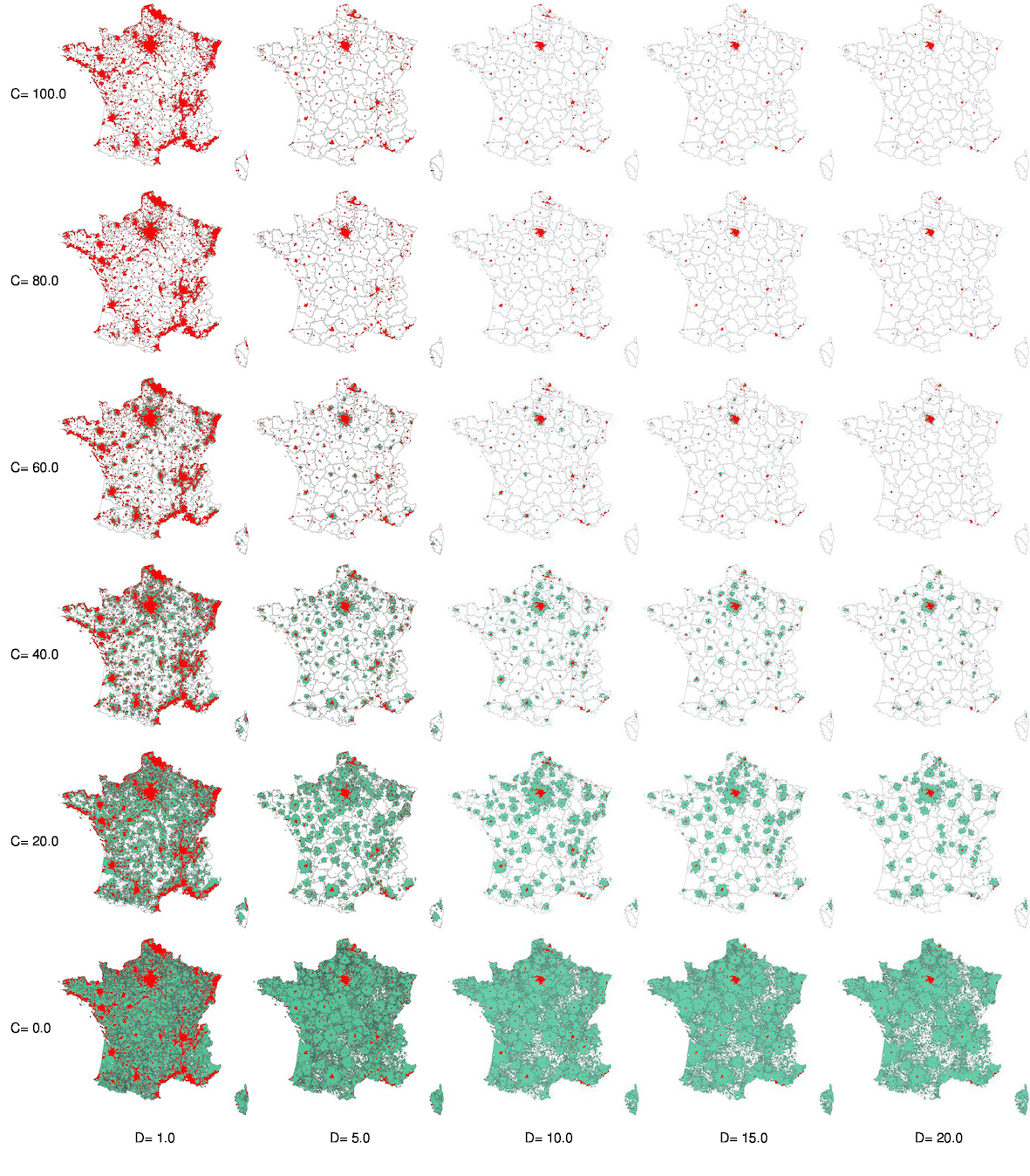}
    \end{center}
    $D$ is the minimum density of residents per ha which defines urban centres (in red). $C$ indicates the share of Commuters (in \%) living in the periphery and working in the density-based urban clusters (in green). $P$, the population minimum, is here set to 0. Aggregation is performed using the 2013 GeoFla geometry of communes. \url{https://www.data.gouv.fr/fr/datasets/geofla-communes/}
\end{figure*}

 In the following sensitivity analysis, we consider 4914 such representations based on the combinations of 39 density values ($D$ from 1 to 20 residents per ha in steps equal to 0.5), 21 commuting cutoffs ($F$ from 0 to 100\% in steps equal to 5) and 6 population cutoffs ($P$ from 0 to 50,000 inhabitants in steps equal to 10,000). Thirty different combinations are shown in figure \ref{fig:defs}. \\

\begin{table*}[t]
\begin{center}
\caption{Three depictions of city definitions (top) and their corresponding urban clusters (bottom). The $R^2$ represents the level of correspondence between each pair and n the number of cities considered.}
\label{table:2}
\begin{tabular}{|c|c|c|}
\hline
\multicolumn{3}{|c|}{City definitions}  \\ \hline
{\it Unit\'{e}s Urbaines}  & {\it Aires Urbaines} & CORINE LandCover \\ \hline
\begin{minipage}{.28\textwidth}
\includegraphics[width=\linewidth]{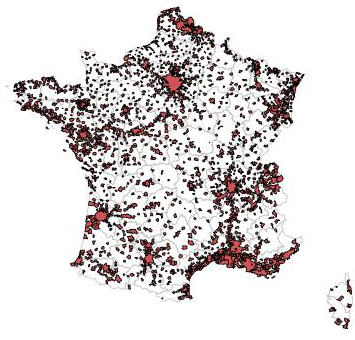}
\end{minipage}
&
\begin{minipage}{.28\textwidth}
\includegraphics[width=\linewidth]{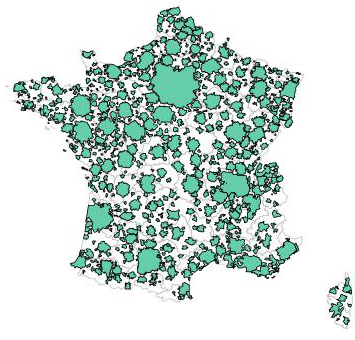}
\end{minipage}
&
\begin{minipage}{.28\textwidth}
\includegraphics[width=\linewidth]{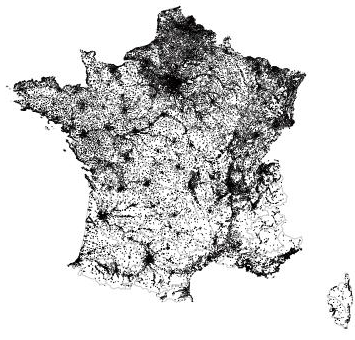}
\end{minipage} \\ 
\hline
\multicolumn{3}{|c|}{Corresponding Urban Clusters} \\ \hline
$D = 1.5, F = 100, P = 0$ & 
$D = 2, F = 35, P = 10,000$ & 
$D = 4.5, F = 100, P = 0$  \\ \hline
\begin{minipage}{.28\textwidth}
\includegraphics[width=\linewidth]{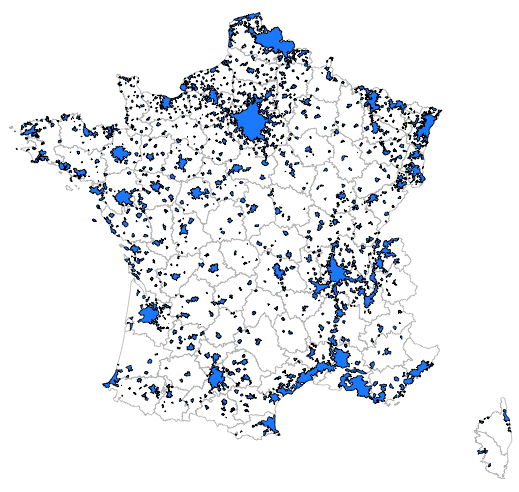}
\end{minipage}
&
\begin{minipage}{.28\textwidth}
\includegraphics[width=\linewidth]{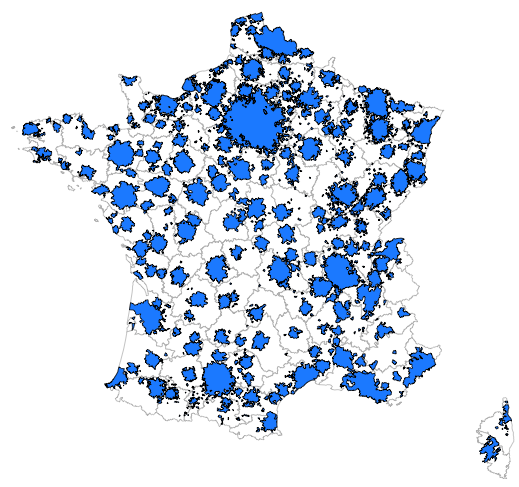}
\end{minipage}
&
\begin{minipage}{.28\textwidth}
\includegraphics[width=\linewidth]{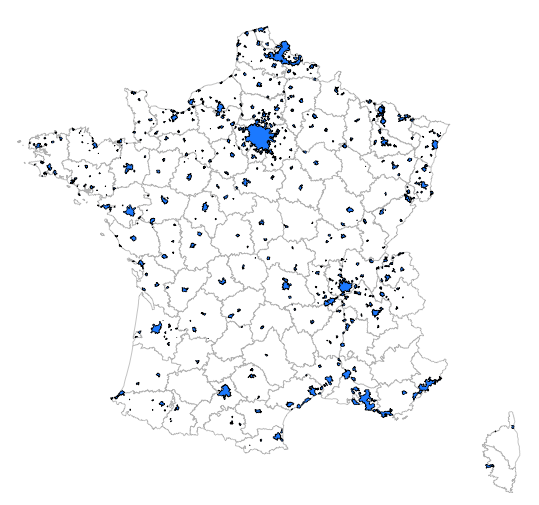}
\end{minipage} \\
\hline
$R^2 = 0.436$  | $n = 1173$ & $R^2 = 0.134$ | $n = 309$ & $R^2 = 0.580$ | $n = 519$ \\ \hline

\end{tabular}
\end{center}
 Source. INSEE: \url{www.insee.fr}, CORINE LandCover 2006 raster data (Urban categories of land use No. 111 and 112): \url{http://www.statistiques.developpement-durable.gouv.fr/donnees-ligne/li/1825.html}
\end{table*} 

 \subsection{Matching with official definitions}
 \label{sec:p13}
 
In order to evaluate the quality of our clusters and their ability to describe the transition between official definitions, we measured the correspondence between each generated cluster definition and three classical urban delineations: Urban Units in 2010, Metropolitan Areas in 2010, and Urbanised Land Use Areas given by CORINE LandCover 2006 raster data (cf. table \ref{table:2} and figure \ref{fig:corrdef} from the supplement). This measure consists of the correlation between urbanised local units in the official and the cluster definitions (a correlation of 1 meaning that both definitions match perfectly: each "urban" local unit by official standards belongs to one of the systematic clusters for a given definitional combination). By doing this, we check if the method is able to generate urban clusters that correspond to the different official definitions, in order to study the transition between them in an almost continuous way. Given the methods used by the statistical office, we expect definitions to differ on the commuting criteria, which should be close to 100\% for the Urban Unit equivalent, and close to 40\% for the Urban Area equivalent, with similar density cutoffs. \\

Indeed, we find a good match between Urban Units and clusters defined as having a density over 1.5 residents per ha, almost no commuting flows ($F = 100\%$) and no population cutoff (table \ref{table:2} and figure \ref{fig:corrdef}, top left). The correlation coefficient between belonging to such a cluster and belonging to an Urban Unit for each commune is 0.66, with an $R^2$ of 44\%. For Urban Areas, we find the following expected values for definitional criteria: a density cutoff of 2.0 (close to that of Urban Units), a commuting cutoff of 35\% (close to the official 40\%) and a larger population cutoff (10,000 persons). However, the match is of weaker quality ($R^2 = 0.134$). \\

The previous definitions correspond to definitions based on administrative boundaries. An alternative but common urban definition comes from the use of CORINE LandCover raster data. When computing the correlation coefficient between the belonging to a cluster and the \% of land classified urban in the CORINE images for each commune, we find a match of better quality ($R^2 = 0.58$) with clusters defined with $D = 4.5$ persons per ha, $F = 100\%$ and no population cutoff.


 \section{Understanding the variations of scaling behaviours}
 \label{sec:p2}
 
 Building on an almost continuous representation of systems of cities, we are able to estimate scaling laws and their sensitivity to the variation of the urban definition criteria. Evaluating the sensitivity of scaling provides an opportunity to identify where scaling regime transitions might take place and to characterise the types of cities to which it corresponds. This will help build models that could account for the emergence of such scaling (and varying) behaviours, based on a non-uniform internal morphology of the cities considered. \\

The variables used to describe urban attributes in the following analysis have been collected at the local level ($ \simeq 36000$ local units in continental France). Socio-economic and travel-to-work data come from the last Population Census and surveys in 2011, whereas land-use data are extracted from CORINE LandCover 2006 raster data, road length derive from an Open Street Map dataset computed by C. Quest in 2014, and housing permits come from governmental open data. A more detailed description is available in the supplementary materials (section S\ref{sec:description}).

  \subsection{Variations of scaling behaviours}
   \label{sec:p21}
   
\begin{figure*}
 \begin{center}
  \caption{The distribution of scaling exponents for selected  attributes over the entire set of city definitions}
    \label{fig:vars}
    \begin{minipage}{\textwidth}
\includegraphics[width=\textwidth]{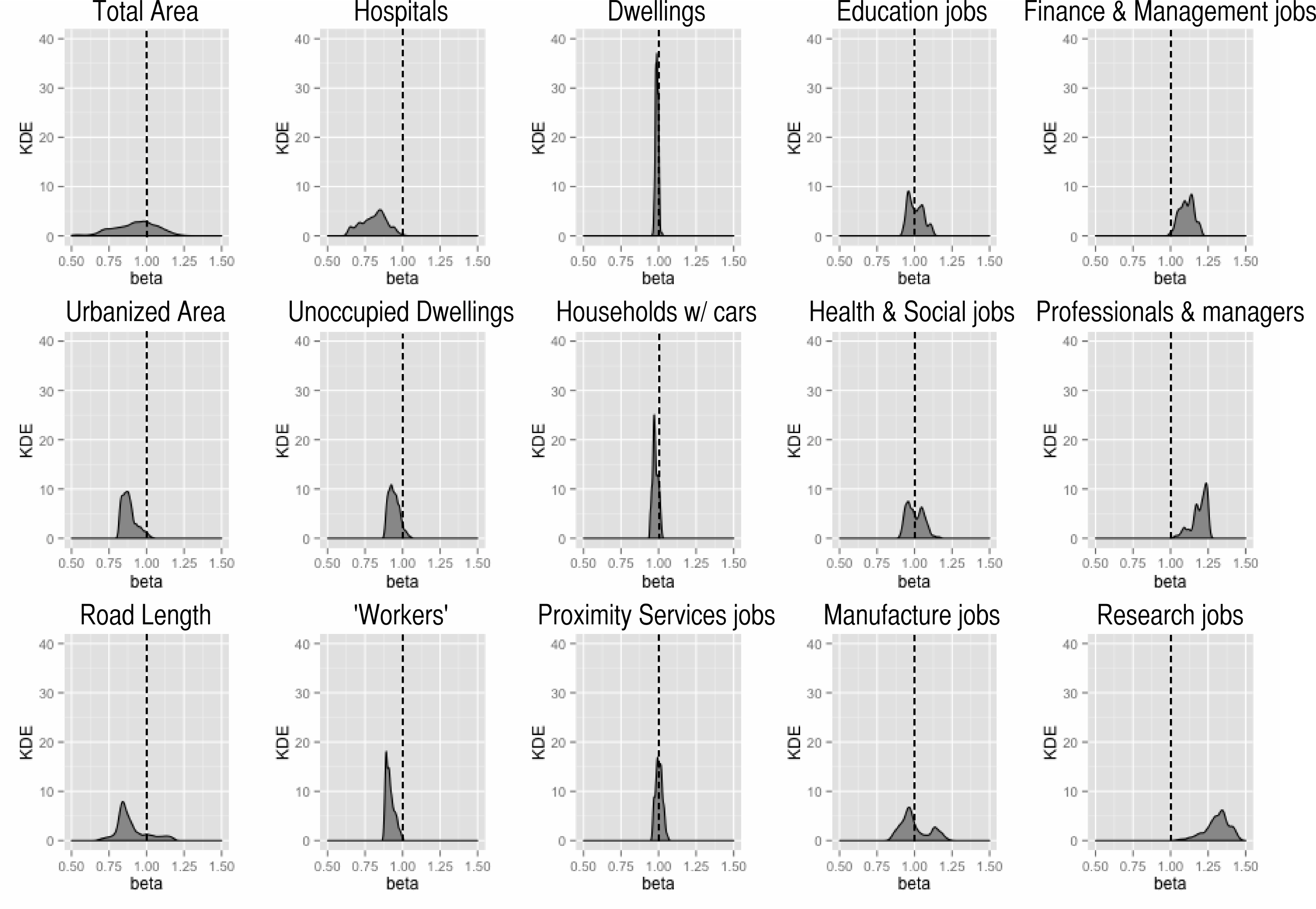}
\end{minipage}
    \end{center}
Kernel Density Estimation (KDE) of scaling exponents of 15 urban attributes from 4914 systems of clusters-cities resulting from the combination of three definition criteria. N.B. : the number of clusters in each regression varies, from 1298 to 48, depending on the restrictive character of the definition criteria
\end{figure*}

   Figure \ref{fig:vars} represents the distribution of the estimated $\beta$ for the 4914 representations of the French urban system. We used a kernel density estimation (\url{http://www.inside-r.org/r-doc/stats/density}) to ease visual comparisons across attributes and cluster definitions. We prefer this representation to histograms in this case because it allows us not to fix any bin number or size, which would have taken different values for each distribution. The kernel procedure, on the contrary, makes the representation of the distributions continuous and comparable. This analysis of density estimations can be summarised by two findings. \\
   
First, we find {\bf substantial variations} in the scaling exponents measured for the different representations of the system of cities. The maximum intervals of estimated exponents often range from sub-linear ($\beta < 1$) to superlinear regimes ($\beta > 1$). For example, the area of urban clusters (figure \ref{fig:vars}, top left) scales from sublinearity ($\beta = 0.33$) to superlinearity ($\beta = 1.29$) with population. The length of paths and roads is the second most volatile variable with respect to urban scaling, as the exponents estimated range from $\beta = 0.66$ to $\beta = 1.20$. These results are not so surprising as the two variables are physical and therefore highly dependent on the spatial definition of cities. However, many other variables (such as the number of jobs in the education or the manufacturing sectors, the number of hospitals, etc.) are also affected by the choice of urban criteria. To summarise, with the exception of the number of dwellings which clearly scales linearly with population ($0.95 < \beta < 1.03$), all the urban attributes considered in the study range across two or more scaling regimes and cannot be classified in a definitive way by a single value for all possible urban definitions.\\

The second finding therefore relates to the {\bf different magnitudes of variations} observed for the different variables under study. If only the number of dwellings is stable over the complete range of city definitions, some attributes appear stable in relation to population for a majority of definitional criteria combinations (in terms of the scaling regime rather than the specific value of the exponent). For instance, the number of households owning a car ($\beta \in [0.94,1.03]$. N.B. this interval corresponds to the amplitude of estimation across definitions, not a confidence interval) and jobs in proximity services ($\beta \in [0.95,1.07]$) scale linearly with population most of the time. Similarly, the number of hospitals, of persons employed as "workers" and the urbanised area scale sublinearly in a majority of representations of the system of cities ($\beta$ belongs respectively to the intervals [0.63;1.02], [0.87;1.01] \& [0.81;1.05]). The number of persons employed as managers and professionals, the jobs in finance or in research are symmetrically mostly superlinear urban attributes ($\beta$ magnitude of respectively [1.02;1.27], [0.98;1.21] \& [0.95;1.5]). Such behaviours are consistent with results obtained with ordinary representations of systems of cities \cite{pumain2006, bettencourt2007}. However, given particular criteria for urban definition, the behaviour of these attributes with city size would change to the opposite scaling regime. Finally, bimodal distributions are clearly noticeable from figure \ref{fig:vars}, on both sides of the linear value where $\beta = 1$ : e.g. the number of jobs in manufacture ([0.83;1.25]), in health and social services ([0.9;1.17]) and in education ([0.92;1.14]). These features suggest that there might be two sets of urban definitional criteria generating two scaling regimes. \\
This observation leads us to inquire much more into the determinants of variations in urban scaling.

   \subsection{What makes urban scaling vary ?}
   \label{sec:p22}

In this section, we look for systematic variations of $\beta$ with the definitional criteria, and ways to interpret and explain them. We proceed in four steps : \\
\textbullet{ Building a typology of cluster definitions and compare their distributions of $\beta$ (section \ref{sec:p221}),} \\
\textbullet{ Using heatmap representations to visualise and compare variations of urban scaling with definitional criteria for the different attributes, producing representations similar to a phase diagram (section \ref{sec:p222}}) \\
\textbullet{ Grouping the most similar heatmaps with hierarchical clustering (section \ref{sec:p223}) and} \\ 
\textbullet{ Comparing extreme estimations with extreme observations from the literature (section \ref{sec:p224}).}
   
\subsubsection{Clusters Typology}
   \label{sec:p221}
     
     Our first approach is to differentiate urban clusters based on their spatial extent and population cutoff. \\
     
          \begin{table*}
\centering
\caption{The number of urban realizations within each typological group}
\label{tab:effectives}
\begin{center}
\begin{tabular}{lclclc|c|}
\hline
Definitions & Population cutoff &No cutoff & Total \\ \hline
Common clusters    & \colorbox{red}{\textcolor{White}{{\bf  630}}}   & \colorbox{BurntOrange}{\textcolor{White}{{\bf  126}}}                & 756                        \\ \hline
Alternative clusters  & \colorbox{black}{\textcolor{White}{{\bf 3465}}}  & \colorbox{Gray}{\textcolor{White}{{\bf  693}}}  & 4158                       \\ \hline
Total                 & 4095                                   & 819                                       & 4914                       \\ \hline
\end{tabular}
\end{center}
Common clusters : $1 \leq D \leq 5$ \& $35 \leq F \leq 100$. \\
Population cutoffs : {10,000; 20,000; 30,000; 40,000; 50,000}.
\end{table*}

 \begin{figure*}
 \begin{center}
  \caption{The distribution of scaling exponents for selected  attributes over the four typologies. Legend : cf. table \ref{tab:effectives}}
\label{fig:realistic}
\includegraphics[width=\textwidth]{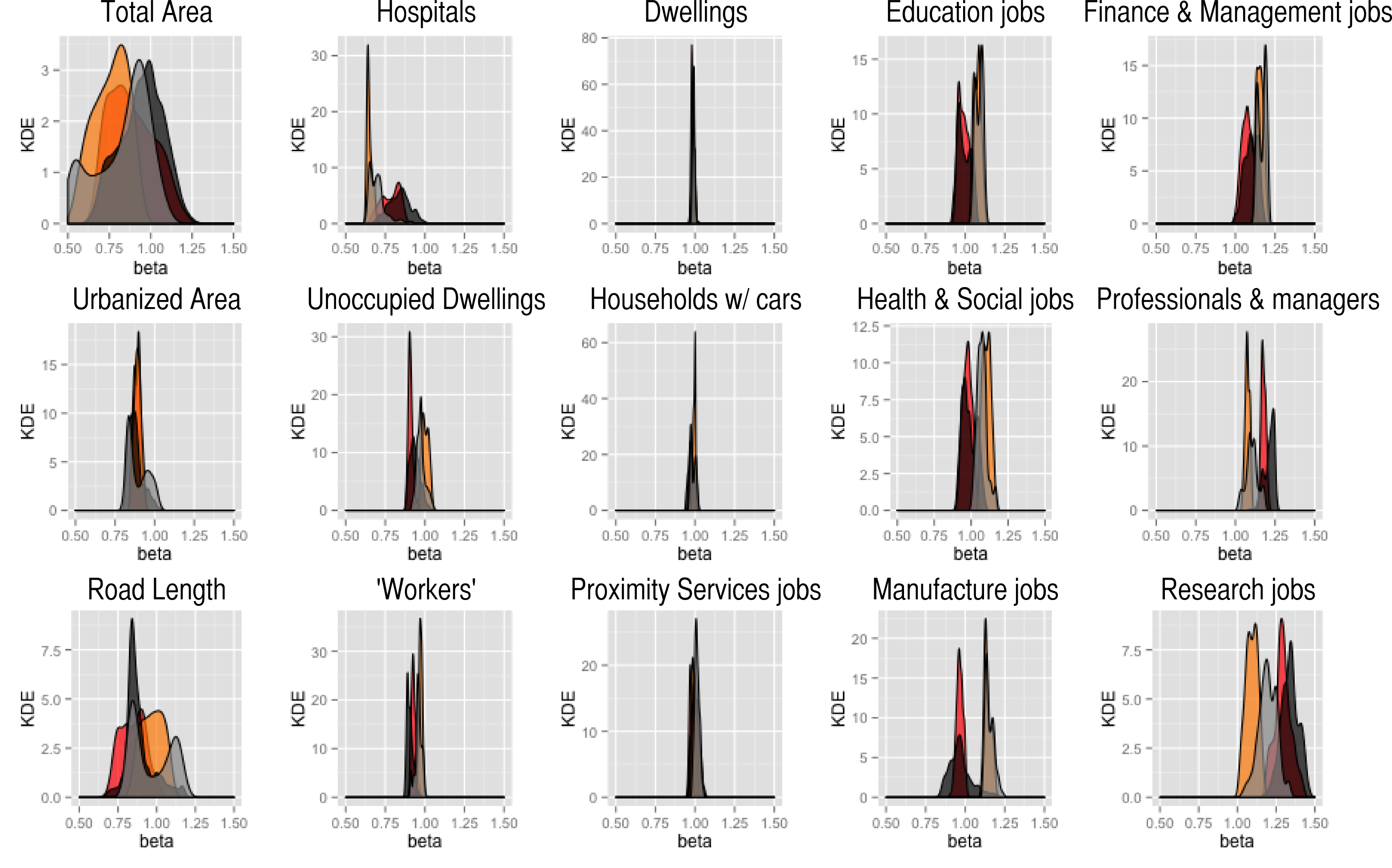}
    \end{center}
\end{figure*}

     We partition our set of urban realisations into two subsets: the first represents the ones which are similar to existing definitions of cities in France and we call them {\bf common clusters} (cf. Supplementary material, figure \ref{fig:corrdef}). {\bf Alternative clusters} represent definitions which deviate from the usual representations of the French system of cities, yet are within reasonable bounds for spatial criteria. 
      Common clusters are defined as centres with population densities ranging from 1 to 5 residents per ha (57\% of Urban Units belong to this interval) and peripheries made of local units from which 35 to 100\% of the active population commutes to the dense centres (the proportion is 40\% for the definition of Metropolitan Areas and virtually 100\% for Urban Units). Alternative clusters represent all remaining definitions at a given population cutoff (cf. figures \ref{fig:defs} and \ref{fig:corrdef}). Population cutoffs do not affect the spatial extension of cities, but drastically change the number of cities considered (the higher the cutoff, the lower their number). For example, with the same criteria of $D = 1.5$ \& $F = 100\%$, there are 1172 urban clusters in France, but less than a third of them have more than 10,000 residents (300), and only 98 meet the $P > 50,000$ definition (for $D = 2$ and $F = 30$, the actual numbers are respectively 991, 341 \& 138). Therefore, among common and alternative clusters, we differentiate between definitions with a population cutoff and definitions without. \\
      
      We find that this first categorisation eases the interpretation of the variations of the scaling exponents, especially in the case of bimodal distributions (figure \ref{fig:realistic}). For example, the two modes in the distribution of the scaling exponents for jobs in the manufacturing sector corresponds to two types of definitions: the ones with a population cutoff and the ones without (independently from the spatial extent of cities, common or alternative). In other words, one finds that manufacturing jobs scale superlinearly with city population when all cities are considered, especially very small aggregates. On the contrary, when one sets a minimal size for cities to be considered, they scale sublinearly to linearly with population (fourth column, figure \ref{fig:realistic}). This result tells us that manufacturing is neither a specialisation of small nor large cities: the bimodal distribution and the transition in scaling regimes could instead indicate either the importance of medium size cities in the concentration of manufacturing jobs or the high threshold in population for economies of scale to appear. The same pattern holds for the education and the health and social sectors, even though the picture is less clear-cut for alternative clusters.  In the case of financial and managerial jobs, the application of a population cutoff does not change the scaling regime but clearly lowers the value of scaling exponent $\beta$, from around 1.2 to around 1.1. This might begin to explain why mixed evidence appears in the literature, depending only on the size of cities considered in the analysis. The opposite holds true for professional and managerial occupations. \\
      
    For some urban indicators (the length and surface variables for instance), the scaling behaviour does not respond monotonically to the application of a population minimum (first column, figure \ref{fig:realistic}). For the majority of others, however, this definitional criterion appears to be the most important to understand variations in the scaling exponents for most urban indicators, more than matters in terms of the spatial extent of cities.

\subsubsection{Scaling heatmaps}
 \label{sec:p222}
            
\begin{figure*}
\begin{center}
\caption{Heatmaps of scaling exponents and goodness of fit across city definitions}
\label{fig:heatmaps}
\includegraphics[width=0.9\textwidth]{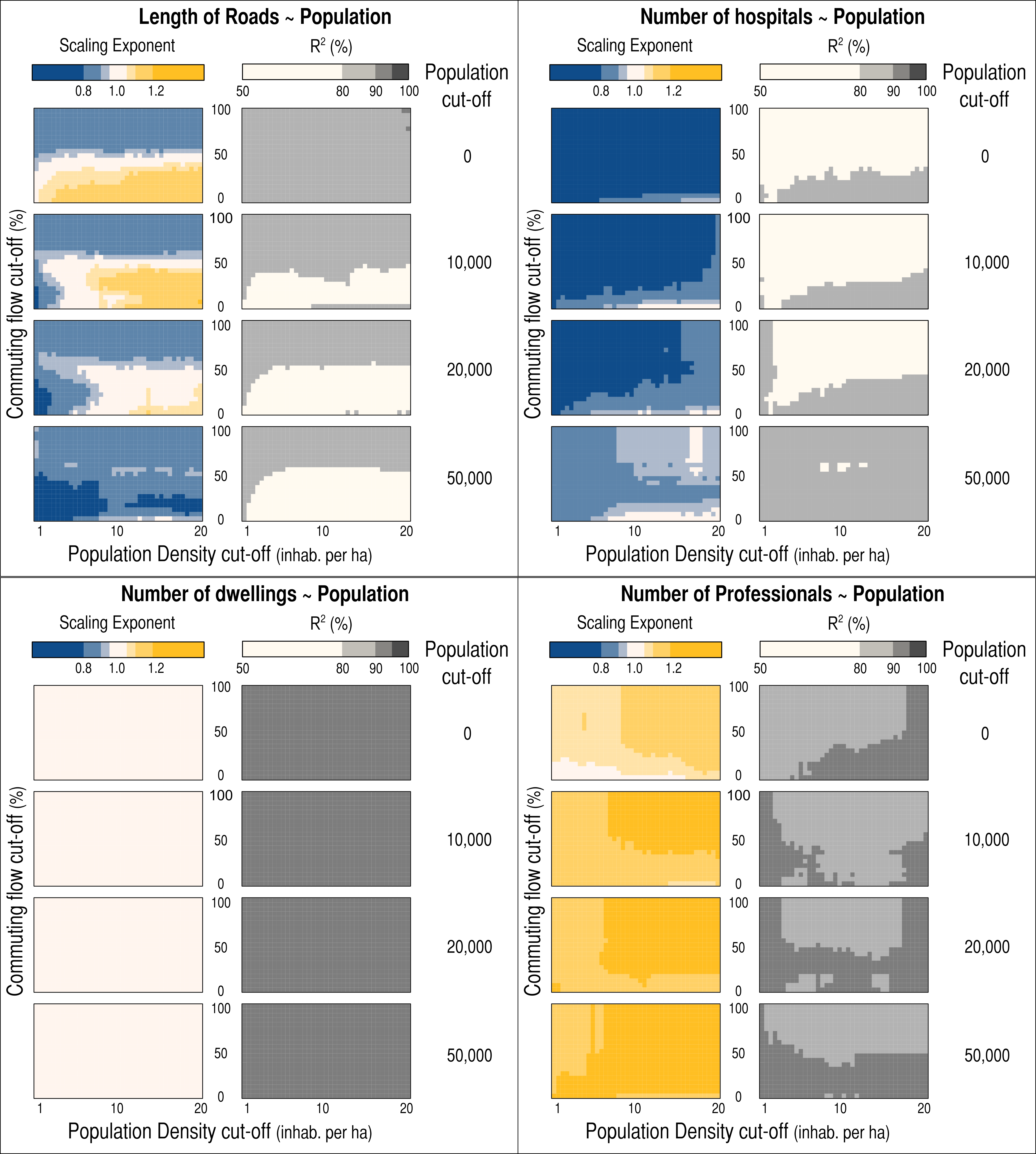}
\end{center}
Each square in the heatmaps represent a scaling regression on all urban clusters resulting from the combination of the three definition criteria-coordinates (density, commuting flows and population cutoff)
\end{figure*}

To provide a more detailed account of the sensitivity of attribute scaling with respect to the definitional criteria and their combination, we use multiple representations of 2D heatmaps. Figure \ref{fig:heatmaps} for instance shows scaling exponents and $R^2$ values for several indicators with population for 3276 definitions.
In the case of the road length (figure \ref{fig:heatmaps} top left), the interplay of the three definitional criteria produces differentiated scaling behaviours. Also, not only do the scaling exponent values vary : the scaling regime (sub- or super-linear) depends on the combination of density, commuting flows and population cutoff. Indeed, although this variable is known to be sublinear from previous reports in the literature (e.g. in \cite{levinson2012}, $\beta = 0.667$, and in \cite{louf2014_sc} $\beta = 0.86$), we show that when relatively dense urban clusters ($>5$ residents per ha) with significant peripheries (flow cutoff $< 50\%$) and low population cutoffs ($<20,000$) are defined, superlinear behaviours are encountered. This behaviour is hidden over a population cutoff of 50,000 inhabitants, but the quality of the regressions also lowers as the population cutoff increases. This means that, in contradiction with some empirical evidence and model predictions, the road length per capita is higher in larger urban aggregates (corresponding to the spatial extents at the bottom right corner of figure \ref{fig:defs}). 
The second heatmap (figure \ref{fig:heatmaps} top right) shows the relatively weak importance of flow and population cutoffs on the estimation of urban scaling for hospitals up to 50,000 persons, and the strong variation due to density over this threshold (towards a linear pattern).
Finally, figure \ref{fig:heatmaps} bottom left and right reveal cases of stability of scaling behaviour associated with robust models : $R^2 > 90\%$ (again, stability here refers to the scaling regime, not the value of the exponent which may vary substantially). The number of dwellings scales linearly with population in any combination of urban definition, whereas the number of professionals and managers scales superlinearly, especially when cities are defined using high density, high commuting integration and high population cutoffs.

\subsubsection{Hierarchical clustering}
 \label{sec:p223}

\begin{figure*}
 \begin{center}
  \caption{Clustering of heatmaps across city definitions}
  \label{fig:CAH}
\includegraphics[width=\textwidth]{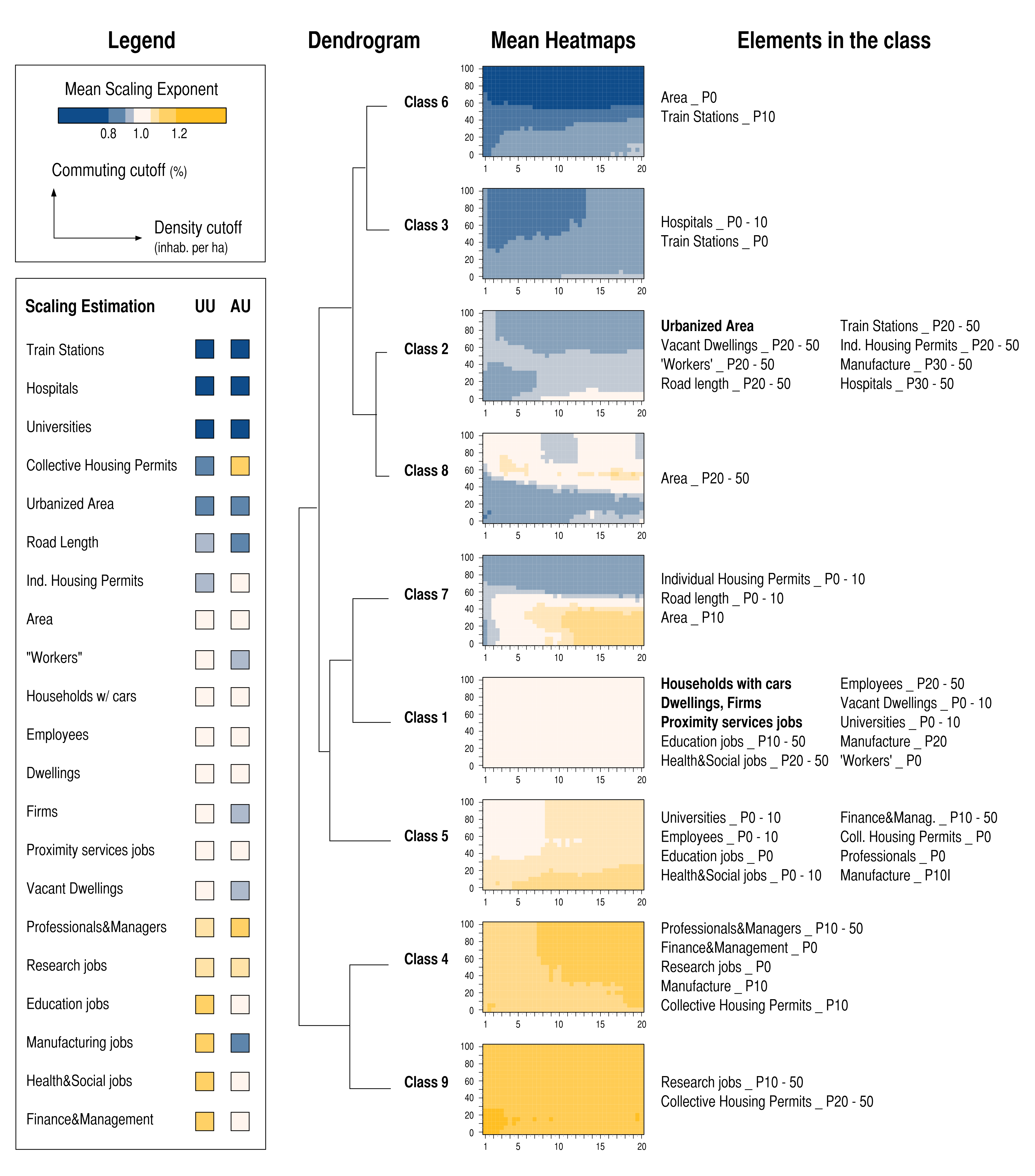}
    \end{center}
    N.B. The "elements in the class" correspond to heatmaps of scaling exponents for each indicator, for each population cutoff. In bold : all the heatmaps for a given indicator belong to the same class. When a given indicator has heatmaps in different classes of the typology depending on the value of the population cutoff, we use the following notation : $P0$ for "no population cutoff", $P10$ for a population cutoff of 10,000 inhabitants, etc.
\end{figure*}

In order to summarise the proximity between values and variations of scaling exponents of 20 different urban attributes, we performed a hierarchical clustering of heatmaps for each population cutoff (figure \ref{fig:CAH}). The $\beta$ estimated on the clusters defined by the combination of 21 density cutoffs and 39 commuter cutoffs represent the variables (819 columns) describing each urban attribute for each population cutoff (the 126 individuals or rows). The clustering procedure therefore distinguishes groups of indicators whose scaling behaviour responds the same way to the city definitional variations. A typology of 9 classes corresponds to clear cuts in the clustering tree (dendrogram) and cover 47.1\% of the total variance. 

\paragraph{Class 1: monotonically linear attributes.}
            
The first class in this typology groups together attributes that scale proportionately with population ($0.95 < \beta < 1.05$) for all values of density and commuting criteria. In the case of the number of dwellings, of firms, of households owning a car and of jobs in the proximity service sector, the scaling exponent measured is linear irrespective of the minimum population cutoff (figure \ref{fig:CAH}). At high population cutoffs (i.e. when only large cities are considered), jobs in the educational, health and social sectors also belong to this class. At low population cutoffs, the number of "workers", manufacturing jobs, vacant dwellings and universities also tend to be linear urban attributes. These results relate to the interpretations of \cite{pumain2006} and \cite{bettencourt2012} concerning indicators of "mature" industries and "basic needs" proportional to urban populations. Following these researchers, linear attributes describe industries mature enough not to require increasing returns to scale to compensate for innovation costs, or goods and services proportionately distributed among city dwellers, since everyone needs a comparable amount in every city of the system (in terms of dwellings, for example). \\

\paragraph{Class 2, 3 \& 6: sub-linear attributes.}
            
In the types 2, 3 \& 6, we find linear to sublinear attributes such as infrastructure, morphological and physical indicators (number of train stations, road length, urbanised area, number of vacant dwellings and hospitals) as well as "obsolete" \cite{pumain2006} industries (manufacturing) that scale overall in the way predicted in the formerly cited theories: they grow slower than proportionately with city populations, revealing economies of scale (for physical attributes) and specialisation of small cities (for obsolete industries). However, these three classes do not appear monotonic in their response to changes in city definitions. For example, the attributes are distributed in various classes depending on the population cutoff applied. Train stations for example scale much more linearly among cities of more than 20,000 inhabitants (class 2) than when cities above 10,000 (class 3) or less (class 6) are taken into account. Indeed, all towns usually contain a local train station in France, leading to a very low exponent associated with no population cutoff (or $P0$, class 6), especially when only urban city cores are considered without their functional periphery (i.e. high commuting cutoffs). Among large cities, potentially those with more than one train station, the behaviour becomes more linear. High sublinearity is therefore more of an artefact due to complete provision in small towns, while the number of stations is indeed almost proportional to the number of city residents. \\

\paragraph{Class 4, 5 \& 9: super-linear attributes.}

As observed in figure \ref{fig:CAH}, research, finance and management jobs, along with professional occupations, collective housing and universities scale linearly to superlinearly, i.e. among classes 4, 5 \& 9. We also find attributes that belong to the superlinear class 5 on the heatmaps at low population cutoffs (Education, health and social jobs, number of employees), while their scaling behaviour under higher cutoffs is monotonically linear (cf. class 1). For these indicators, the more restrictive the delineation of city centres (i.e. small periphery and large population), the more linearly they grow with population. Therefore, the superlinear scaling observed in class 5 for weak flows and low population cutoffs reveals a higher provision in public services and employment opportunities in urban spaces compared to rural ones, rather than a size differentiation among cities. \\

\paragraph{Class 7 \& 8: mixed attributes.}
            
Classes 7 \& 8 represent non-monotonic regimes of urban scaling (of particular interest to this paper). Class 7 groups heatmaps for road length, area and individual housing at low population cutoffs, and show a combination of three regimes: sublinear (for commuting flows $F >50\%$); linear (for $F < 50\%$ \& $D < 10$ persons per ha); superlinear ($F < 50\% $ \&  $D > 10$). When we refer to figure \ref{fig:defs}, this means that those indicators of suburban morphology grow less than proportionately with population in city centres (top half of the figure), proportionately in vast configuration of cities that represent most of the French territory (bottom left), and more than proportionately with population when cities are considered as high density kernels surrounded by large peripheries (bottom right). This obvious conclusion highlights the fact that in the case of physical urban attributes, the estimation of scaling exponents is directly related to the choice of city definition.
Class 8 comprises only area heatmaps for high population cutoffs and is to a lesser extent affected by the density criterion to define city centres. The resulting picture is somehow opposite to class 7 where area scales linearly to superlinearly in city centres (high $F$) and sublinearly in systems with vast peripheries (low $F$). In other words: large metropolitan areas get denser as they grow, while the surface occupied by city centres is more or less proportional to population (probably revealing areas occupied by offices and large infrastructures compensating for higher residential density). \\

An interesting feature of these mixed classes is the fact that the transition between the two regimes happens around the flow cutoff value of 40\%, the one chosen by INSEE to define peripheries of metropolitan areas (AU). This cutoff corresponds mostly to linear behaviour in the classes, but also to a transition space between two radically different scaling regimes. The linear scaling at this value therefore hides high variability when cutoffs are slightly pertubed. It is thus not robust to city definition, indicating that either 40\% is "the true value for cities", or more probably that there is no interpretation of the scaling of these variables independently from the criteria used to define cities. \\

\paragraph{Several types at different population cutoffs: manufacturing jobs.}

A special case is that of manufacture. The scaling heatmaps for this indicator are found in linear, sub- and superlinear classes according to which population cutoff is considered. Scaling laws are supposed to account for regular variations across several orders of magnitude, and specifically model the distribution of attributes of cities of different population sizes. These results for manufacture therefore suggest that the power law adjustment might not be the most interesting one to describe the evolution of manufacturing jobs with city population. It also suggests that other factors of explanation are necessary to understand the distribution of manufacturing jobs in French cities, such as resource and path-dependency, regional particularities and economic cycles (cf. section \ref{sec:p3} and residual analysis from section S\ref{sec:residuals} in supplement). \\
 
\paragraph{Clustering summary.}

The clustering of heatmaps has thus provided a synthetic way to describe the dominant regime and variation of scaling measures of 20 indicators with respect to urban definitional criteria. The amplitude of variations appears marginal for many indicators for which the dominant scaling regime corresponds to the one predicted theoretically or from ordinary definitions of cities. We also identified groups of urban attributes for which the variation of the scaling exponent depends quantitatively on density and flow cutoffs. Physical attributes for instance are not independent from the spatial definitions of cities. There, scaling behaviour appears  impossible to characterise independently from city definitional criteria. On the other hand, we found that social and public services seem to exhibit constant returns to scale over a minimum threshold of population. Finally, manufacturing jobs have been found to be loosely linked to population and are better described by other urban features (especially with respect to their history of early diffusion in north-eastern France).\\

A last attempt at understanding and explaining variations in scaling comes from the confrontation of extreme values recorded in the literature and by our methodology.

\subsubsection{Extreme Scaling}
\label{sec:p224}
                     
This last section examines the extreme cases of scaling measured on systems of cities: in the literature and with a systematic definition. Table \ref{table:4} gives the maximum intervals of scaling exponents found in the literature and in this study among the 4914 combinations of definitional criteria. \\

 The total area of cities is an interesting example of such extreme variations reported in the literature as well as in our own study. The lowest scaling exponents measured are very low ($\beta = 0.3$) and represent extremely sublinear behaviour compared to the minimum values we found in the literature (0.676 in \cite{batty2011}). In the case of contemporary urban France, we found this minimum value for a very restrictive definition of cities corresponding to narrow centres without peripheries ($D = 13.5, F = 100, P = 0 | N = 202$). On the contrary, a superlinear regime ($\beta = 1.2$) comparable to that found by Veregin and Tobler for 366 US cities in 1980 corresponds to a relatively common definition of metropolitan areas ($D = 7.5, F = 45, P = 10,000 | N = 200$). In this case, the dramatic variation of scaling estimation depends on the way cities are spatially defined, as they mechanistically affect the ratio of surface per inhabitant. \\

For finance and lifestyle measures such as the urbanised area and the number of households with a car, the maximum variations happen with respect to the way the periphery is taken into account. This seems consistent with how morphology and lifestyle interact: suburban spaces are dependent on the use of car, and therefore this variable is the closest to superlinearity in a definition of cities that comprises large peripheries ($\beta = 1.03 | D = 20, F = 5, P = 10,000 | N = 98$) and the closest to sublinearity in narrow centres with almost no commuting ($\beta = 0.94 | D = 17.5, F = 95, P = 50,000 | N = 50$). The interesting insight provided here by looking at the scaling behaviour is that the low consumption of cars and housing per capita in city centres is reinforced as they grow in size. On the contrary, peripheries tend to exacerbate suburban lifestyles when they belong to larger cities. \\ 

In the case of research and education, the distinction between min and max $\beta$ is associated with the density dimension (the way city centres are defined). Those urban attributes are known to be strongly linked to the density of interactions \cite{bettencourt2007, arbesman2011}. What this analysis brings is an insight into extreme behaviours not reported so far in the literature (such as superlinear scaling for roads and sublinear scaling for research under extensive definitions probably comprising rural spaces). \\

The number of jobs in proximity services appears different from the previous category as the minimum and maximum scaling values are opposed in the way the ratio of centre and periphery is defined. Indeed, this variable scales sublinearly when clusters correspond to large dense centres with a restricted periphery ($\beta = 0.95 | D = 5, F = 55, P = 20,000 | N = 163$) and superlinearly when city cores are very narrow and peripheries extensive ($\beta = 1.07 | D = 16, F = 0, P = 10,000 | N = 142$). This could mean that, although basic services are subject to economies of scale in the largest cities, they are not equally profitable in suburban spaces and are provided more systematically in the peripheries of large cities rather than in the periphery of smaller ones. \\

Finally, health and social services jobs show extreme scaling behaviour under similar spatial definitions of cities, but at different population cutoffs. This urban attribute is the most sublinear at the top of the urban hierarchy ($\beta = 0.90 | D = 4, F = 65, P = 50,000 | N = 88$) and most superlinear when the entire spectrum of city size is considered ($\beta = 1.17 | D = 1.5, F = 95, P = 0 | N = 1094$). This might suggest the existence of a critical size to provide such services and their subsequent economies of scale.  \\


\section{Conclusion : from quantitative to qualitative changes in the urban hierarchy ?}
 \label{sec:p3}

\textquote{{\it Whether a particular class of prosocial behavior scales linearly, superlinearly, or sublinearly might be dependent on two factors: the locality of one's interactions, and feedback of interaction. For locality, this refers to whether or not you are limited to the individuals around you, or can interact with anyone in the city. [...] The potential for understanding the spectrum of scaling of prosocial behaviours points to the wide variety of aspects of seemingly related behaviours. Their implications for urban growth are intriguing and merit further examination}} \cite[p.2158]{arbesman2011} \\

Scaling laws were first proposed as an interesting tool to summarise distributions of characteristics in a system of cities with respect to city size over several orders of magnitude. It lies on the conception that cities share common attributes across a wide size spectrum. This proposition finds roots in the long quest to find what makes cities identifiable (specifically urban features). Social scientists agree on the fact that we recognise cities by their function as social interactions maximisers. They do so by {\bf concentrating} a larger number of {\bf heterogeneous social agents} in a limited space (hence, {\bf dense}). Those cities take part in a system of cities at a larger scale, characterised by a regular hierarchy of sizes, competition (in space resulting in regular spatial patterns of settlements) and cooperation (resulting in complementary profiles of economic and functional specialisation) \cite{pumain2006}. Therefore, the analysis of systems of cities should not be restricted to large cities only (a constraint generally imposed by data), as small cities are an essential part of urban systems. \\

\begin{figure*}
 \begin{center}
\caption{Variation of urban scaling with population cutoff for official definitions}
\label{figure:officialandcutoff}
\includegraphics[width=.8\textwidth]{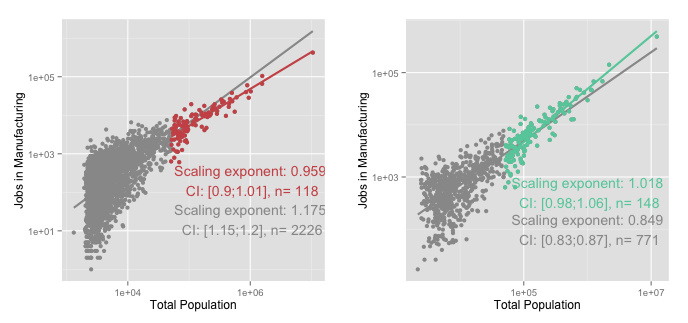}
\includegraphics[width=.8\textwidth]{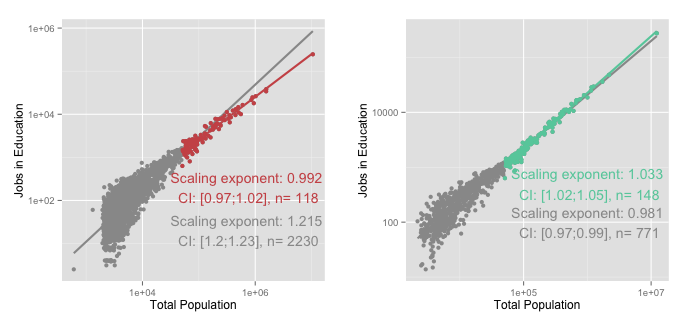}
     \end{center}
Left : UUs. Right : AUs. In color : Over 50,000 residents.
\end{figure*}

Regimes of sub- and superlinearity indicate quantitative changes which occur with size variations. Moreover, several results from this paper indicate the existence of high variability of scaling exponents with respect to variations in city definition. The transition from one scaling regime to another when population cutoffs vary was found to be the most important criterion in most cases (and the easiest to harmonise in comparative studies). For example, if we go back to the paradoxes implied in table \ref{table:1}, we find that scaling behaviours with the two official definitions start to match when a population cutoff is applied (cf. figure \ref{figure:officialandcutoff}). Over 50,000 residents, city centres behave similarly to metropolitan areas with respect to socio-economic criteria. The paradox appears mostly due to the large dispersion encountered among small cities (cf. figure \ref{figure:officialandcutoff}). These deviations from the power law adjustment are neither random errors nor systematic biases for particular cities, for they reveal factors unrelated to city size that play an important role in the explanation of the attribute location, for example: coastal accessibility for secondary houses, regional diffusion in manufacturing, etc. (cf. residual maps from figure \ref{fig:residualsReg} in supplement).\\

\begin{table*}
\begin{center}
\caption{Maximum scaling registered in the literature compared to estimated values with a systematic definition of French cities}
\label{table:4}
\begin{tabular}{|c|c|c|c|c|c|}
    \hline
Urban attribute	& Max. interval for clusters & $\beta_{min}$ in literature [ref] & $\beta_{max}$ in literature [ref] & $\beta_{UU}$ & $\beta_{AU}$\\
    \hline
    Total Area & [0.334; 1.291] & 0.676 \cite{batty2011} & 1.163 \cite{veregin1997} & 0.959 & 0.995 \\
    \hline
     Road Length & [0.664; 1.202] & 0.667 \cite{levinson2012} & 0.86 \cite{louf2014_sc} & 0.903 & 0.888 \\
    \hline
     Research & [0.951; 1.496] & 1.174 \cite{bettencourt2007} & 1.67 \cite{pumain2006} & 1.094 & 1.079 \\
    \hline
     Health and Social & [0.898; 1.171] & 0.95 \cite{pumain2006} & 0.98 \cite{pumain2006} & 1.136 & 1.013 \\
    \hline
\end{tabular}
\end{center}
For more values from the literature and the corresponding definition of cities (territory, date, delineation), cf. Supplement
\end{table*} 

The main finding of this paper is that urban scaling is relative to the definition of cities, and most importantly that variations with respect to definitional criteria are neither random (since residuals can be interpreted on a case by case basis) nor universal for all the variables under study. Instead, some attributes are more sensitive to the spatial delineation of centres and peripheries (interaction-based activity, life-style attributes), while others respond to changes in population cutoffs (for instance: infrastructures with high fixed costs). Although there seems to be no single "good" definition to study urban scaling, what we found is that while selecting one for comparison in time or between national systems, this choice should depend on the attribute under consideration. In particular, the minimum population used to identify cities plays a major role in the scaling variation, and should be integrated fully in the interpretation of results. \\

As a limitation to this work, we could point to the fact that the assumption that cities are monocentric (by attaching commuters to a single centre) could be refined to provide results more in accordance with the diversity of urban forms (and in particular polycentricity, cf. \cite{lenechet2015}). Also, according to Guilluy \cite{guilluy2013}, the "fracture" between central metropolises and peripheral suburbs and rural spaces in France is recent and increasing. It favours the concentration of extremes categories of workers and social classes in globalised metropolises and rejects low and middle social classes to a peripheral France of suburbs and small cities. Provided the collection of temporal data, our method should allow to test these assumptions quantitatively by showing larger variations of urban scaling with city definitions over time for economic and sociologic categories. \\

With this insight and a deeper search for processes linking intra-urban features to interurban scaling, a logical continuation of this work would be to build models able to simulate empirical scaling patterns. These models (i.e. models that seek to explain the role played by city size in the distribution of functions and infrastructures) should now account for the contrasted behaviours of the different attributes, especially at the intra-urban scale. That is, if absolute values of urban scaling are meaningless, there is no point in trying to validate models against them. Instead, the validation goal of generative models should be to reproduce variations in scaling with respect to the variations of the definitional criteria. \\



\end{multicols}

\newpage

\begin{center}
\textbf{\LARGE Supplementary Materials}
\end{center}
\setcounter{equation}{0}
\setcounter{figure}{0}
\setcounter{table}{0}
\setcounter{section}{0}
\makeatletter
\renewcommand{\theequation}{S\arabic{equation}}
\renewcommand{\thefigure}{S\arabic{figure}}
\renewcommand{\thetable}{S\arabic{table}}
  
  \section{Correlation between systematic clusters and official definitions} 
\label{sec:corrdefs}

\begin{figure}[h!]
 \begin{center}
\caption{Correspondence between clusters definitions and official cities}
\label{fig:corrdef}
\includegraphics[width=0.7\textwidth]{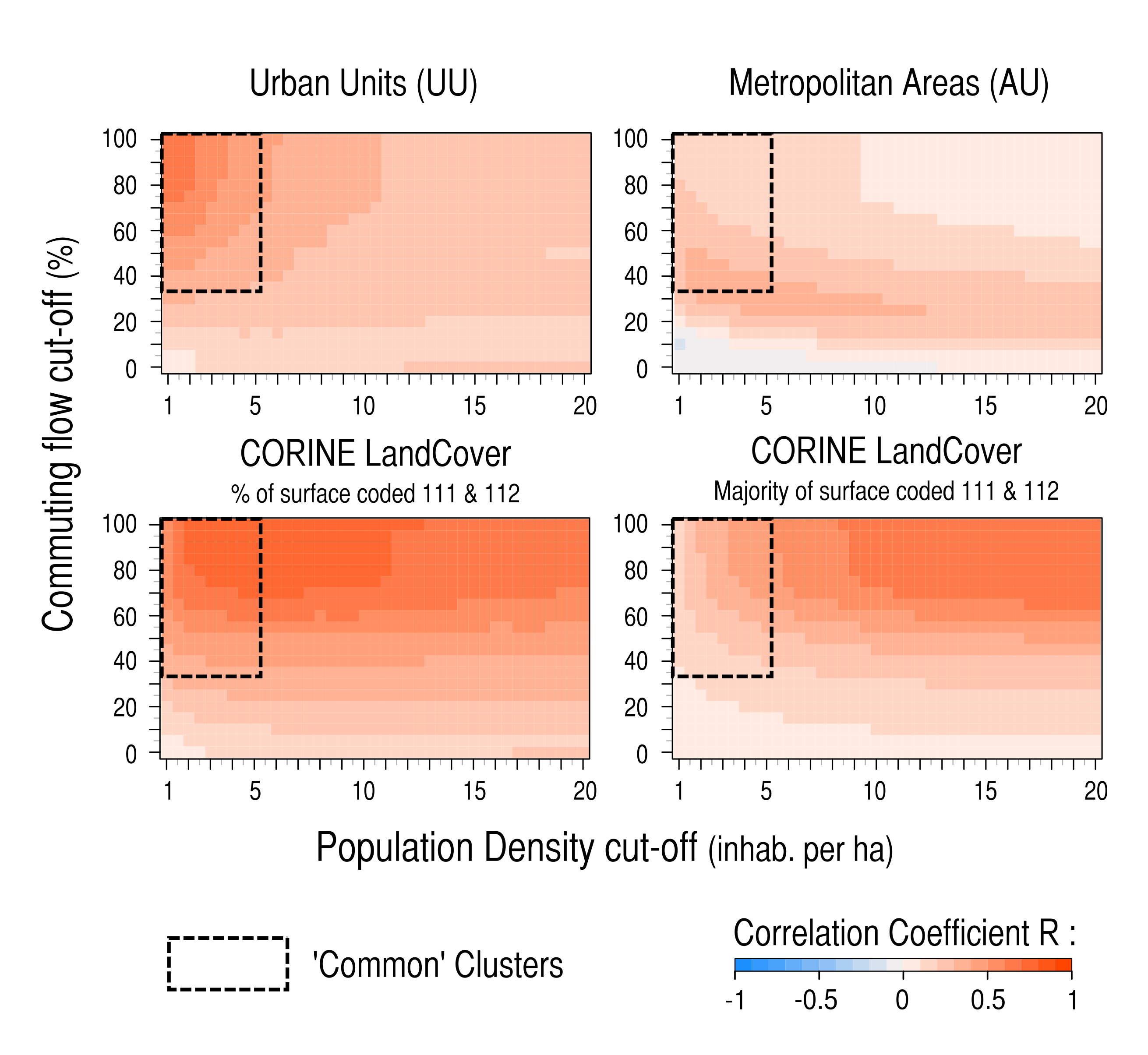}
    \end{center}
\end{figure}

\newpage

\section{Description and sources of urban attributes used in the paper} 
\label{sec:description}

\begin{table}[h!]
\centering
\caption{Sources of data used in the paper}
\label{my-label}
\begin{tabular}{lclclc|c|}
\hline
Type of data & Name & Description  & Source  \\ \hline
\multirow{3}{*}{Geography} 	& 	GeoFla 2013 					& Shape Files 		& 	www.data.gouv.fr  \\
								&	CORINE LandCover 		& Land use 		& www.statistiques.developpement-durable.gouv.fr  \\
								&	MOBPRO 2011 		& Commuting flows	& 	www.insee.fr \\ \hline
\multirow{7}{*}{Attributes} & base-cc-emploi-pop-active-2011 			& households etc.  	& 	www.insee.fr  \\					
							& CLAP 2011 					& Jobs by sector	& 	www.insee.fr  \\
							& equip-tour-transp 2013 			& Infrastructures	& 	www.insee.fr   \\
							& equip-serv-sante 2013 			& Hospitals		& 	www.insee.fr   \\
							& equip-serv-ens-sup-form-serv 2013 & Universities		& 	www.insee.fr  \\
							& OpenStreetMap 2014 			& Length of roads* 	& 	www.data.gouv.fr   \\
							& Sit@del2 2011 				& Housing permits 	& www.data.gouv.fr\\ \hline

\end{tabular}
*primary, secondary, tertiary, motorway and trunk categories added together
\end{table}

Note: The term "worker" here stands for the French category {\it ouvrier}, that is a social occupation defined by the census, which corresponds partially to a plant or manual worker.
Proximity services relate to jobs in everyday services except distribution, transportation, education and health. For example, they include jobs such as hairdressers or laundry services.
Source: \url{http://www.insee.fr/fr/themes/detail.asp?reg_id=99&ref_id=analyse}.
Housing permits are counted as the ones authorised in 2011. Source: \url{https://www.data.gouv.fr/fr/datasets/permis-de-construire-pc-permis-d-amenager-pa-et-declaration-prealable-dp-sit-del2/} \\
Paris, Lyon and Marseille are here considered as single {\it communes} (i.e. not disaggregated into smaller units known as {\it arrondissements}).

\newpage

\section{Residuals from scaling regression} 
\label{sec:residuals}

 \begin{figure}[h!]
 \begin{center}
\caption{Residuals of regression with population (in logs)}
\label{fig:residualsReg}
 \begin{minipage}[b]{\textwidth}
\includegraphics[width=.3\textwidth]{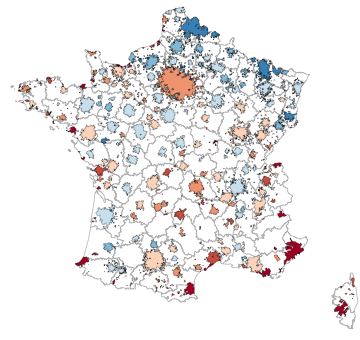}
\includegraphics[width=.3\textwidth]{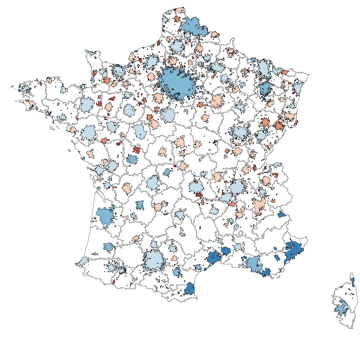}
\includegraphics[width=.3\textwidth]{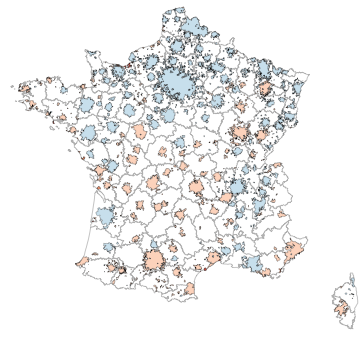}
    \end{minipage}
    \end{center}
    From left to right: Residuals in the Number of Unoccupied Secondary Dwellings, Number of Jobs in Manufacturing, Number of Dwellings. Residuals are obtained from a regression in logs. Red values correspond to a value for the attribute higher in reality than expected with respect the size of the city, blue means that scaling over-estimates the value for the attribute based on the size only. Bright colours indicate large deviations from the scaling estimation.\\
    Cluster Definition: $D = 4, F = 40, P = 0$.
\end{figure}

\newpage

\section{Scaling and allometry form the literature} 
\label{sec:literature}

\begin{table}[h!]
\centering
\caption{Diversity of urban scaling exponents from the literature}
\label{my-label}
\begin{tabular}{lclclclclclclclcl}
\hline
VARIABLE & $\beta$ & CI (95\%)  & $R^2$    & DATE & COUNTRY      & URBAN DEFINITION               & N    & REF      \\ \hline
\multirow{8}{*}{TotalArea}        & 1.163 &                 & 0.774 & 1980 & USA          & "cities"                       & 366  & \cite{sveregin1997}  \\
                                  & 1.043 &                 & 0.903 & 1990 & East Anglia  & "cities"                       & 70   & \cite{sbatty1994}  \\
                                  & 0.808 &                 & 0.756 & 1990 & SouthEast UK & "cities"                       & 801  & \cite{sbatty1994}  \\
                                  & 1.014 &                 & 0.76  & 2001 & Europe       & "cities"                       & 386  & \cite{sfuller2009} \\
                                  & 0.946 &                 &       & 2001 & UK           & "cities"                       & 67   &\cite{sfuller2009}   \\
                                  & 0.765 &                 & 0.637 & 2001 & UK           & metropolitan local authorities & 100  & \cite{sbatty2011}  \\
                                  & 0.676 &                 & 0.309 & 2005 & USA          & SMSA                           & 355  & \cite{sbatty2011}  \\
                                  & 0.85  & {[}0.84;0.86{]} & 0.93  & 2010 & USA          & Urban Areas                    & 3540 &\cite{slouf2014_sc} \\ \hline
\multirow{3}{*}{Road Length}      & 0.86  & {[}0.84;0.88{]} & 0.92  & 2011 & USA          & Urban Areas                    & 441  & \cite{slouf2014_sc}   \\
                                  & 0.667 &                 & 0.65  & 2010 & USA          & MSA                            & 50   & \cite{slevinson2012}  \\
                                  & 0.849 & {[}0.81;0.89{]} & 0.65  & 2006 & USA          & metropolitan areas             &      & \cite{sbettencourt2012}  \\ \hline
\multirow{5}{*}{Research}         & 1.211 &                 & 0.63  & 1987 & USA          & MSA                            & 227  & \cite{sbettencourt2007}  \\
                                  & 1.174 &                 & 0.67  & 1997 & USA          & MSA                            & 266  &\cite{sbettencourt2007}  \\
                                  & 1.185 &                 & 0.69  & 2002 & USA          & MSA                            & 278  & \cite{sbettencourt2007} \\
                                  & 1.54  &                 &       & 2000 & USA          & SMAs                           & 331  & \cite{spumain2006} \\
                                  & 1.67  & {[}1.54;1.80{]} & 0.64  & 1999 & France       & Aires urbaines                 & 350  & \cite{spumain2006} \\ \hline
\multirow{2}{*}{Health \& Social} & 0.95  &                 &       & 2000 & USA          & SMAs                           & 331  & \cite{spumain2006} \\
                                  & 0.98  & {[}0.94;1.02{]} & 0.89  & 1999 & France       & Aires urbaines                 & 350  &\cite{spumain2006} \\ 
\hline
\end{tabular}
\end{table}


\begin{thebibliography}{99} 
   
\bibitem{arbesman2011}
Arbesman S., Christakis N. A., 2011, Scaling of prosocial behavior in cities
\newblock Physica A, 390, 2155:2159

\bibitem{arcaute2015}
Arcaute, E., Hatna, E., Ferguson, P., Youn, H., Johansson, A., Batty, M., 2015, Constructing cities, deconstructing scaling laws
\newblock Journal of The Royal Society Interface, 12(102), 20140745.

\bibitem{batty2011}
Batty, M., Ferguson, P., 2011, Defining city size. 
\newblock Environment and Planning B: Planning and Design, 38(5), 753-756.

\bibitem{berry1964}
Berry B. J. L., 1964, Cities as systems within systems of cities
\newblock Papers in Regional Science, 13(1), 149:163.

\bibitem{bettencourt2007}
Bettencourt, L. M., Lobo, J., Strumsky, D., 2007, Invention in the city: Increasing returns to patenting as a scaling function of metropolitan size
\newblock Research Policy, 36(1), 107:120.

\bibitem{bettencourt2012}
Bettencourt L., 2012, The origins of scaling in cities
\newblock Science, 340, 1438:1441

\bibitem{cristelli2012}
Cristelli, M., Batty, M., Pietronero, L. (2012). There is More than a Power Law in Zipf. 
\newblock Scientific Reports, 2. 

\bibitem{guerin1995}
Gu\'{e}rin, F., (1995). Rank-size distribution and the process of urban growth. 
\newblock Urban studies, 32(3), 551:562.

\bibitem{guerois2002}
Gu\'{e}rois, M., Paulus, F., 2002, Commune Centre, Agglom\'{e}ration, Aire Urbaine : Quelle Pertinence Pour L'\'{e}tude Des Villes?
\newblock Cybergeo, 212 : 15 

\bibitem{guilluy2013}
Guilluy C, 2013, Fractures fran\c caises, Paris: Flammarion, Champs essais, 186p.

\bibitem{kuhnert2006}
Kuhnert C., Helbing D., West G., 2006, Scaling laws in urban supply networks
\newblock Physica A, 363, 96:103

\bibitem{lenechet2015}
Le N\'{e}chet, F. (2015). De la forme urbaine \`{a} la structure m\'{e}tropolitaine: une typologie de la configuration interne des densit\'{e}s pour les principales m\'{e}tropoles europ\'{e}ennes de l'Audit Urbain. 
\newblock Cybergeo: European Journal of Geography, document 709

\bibitem{levinson2012}
Levinson D., 2012, Network Structure and City size
\newblock PLoS ONE, 7(1).

\bibitem{lobo2013}
Lobo, J., Bettencourt, L. M., Strumsky, D., West, G. B., 2013, Urban scaling and the production function for cities
\newblock PloS one, 8(3), e58407

\bibitem{louf2014_epb}
Louf R., Barthelemy M., 2014a, Commentary. Scaling : lost in the smog
\newblock Environment and Planning B, vol. 41, pp. 767:769.

\bibitem{louf2014_sc}
Louf R., Barthelemy M., 2014b, How congestion shapes cities : from mobility patterns to scaling
\newblock Scientific reports, Vol.4, No. 5561.

\bibitem{malecki1980}
Malecki, E. J., 1980, Growth and change in the analysis of rank-size distributions: empirical findings
\newblock Environment and Planning A, 12(1), 41:52.

\bibitem{ortman2014}
Ortman, S. G., Cabaniss, A. H., Sturm, J. O., Bettencourt, L. M., 2014, The pre-history of urban scaling
\newblock PloS one, 9(2), e87902.

\bibitem{parr2007}
Parr, J. B., 2007, Spatial definitions of the city: four perspectives. 
\newblock Urban Studies, 44(2), 381:392.

\bibitem{pumain2006}
Pumain, D., Paulus, F., Vacchiani-Marcuzzo, C., Lobo, J., 2006, An evolutionary theory for interpreting urban scaling laws
\newblock Cybergeo: European Journal of Geography, document 343.

\bibitem{shalizi2001}
Shalizi C. R., 2001, Scaling and hierarchy in urban economies
\newblock arXiv preprint, arXiv:1102.4101v2

\bibitem{veregin1997}
Veregin, H., Tobler, W. R., 1997, Allometric relationships in the structure of street-level databases. 
\newblock Computers, environment and urban systems, 21(3), 277-290

\end{thebibliography}

\begin{thebibliography}{11}
\bibitem{sbatty2011}
Batty, M., Ferguson, P., 2011, Defining city size. 
\newblock Environment and Planning B: Planning and Design, 38(5), 753-756.

\bibitem{sbatty1994}
Batty, M., Longley, P. A., 1994, Fractal cities: a geometry of form and function. 
\newblock Academic Press.

\bibitem{sbettencourt2007}
Bettencourt, L. M., Lobo, J., Strumsky, D., 2007, Invention in the city: Increasing returns to patenting as a scaling function of metropolitan size
\newblock Research Policy, 36(1), 107:120.

\bibitem{sbettencourt2012}
Bettencourt L., 2012, The origins of scaling in cities
\newblock Science, vol. 340, pp 1438:1441

\bibitem{sfuller2009}
Fuller, R. A., Gaston, K. J., 2009, The scaling of green space coverage in European cities. 
\newblock Biology letters, 5(3), 352-355.

\bibitem{slevinson2012}
Levinson D., 2012, Network Structure and City size
\newblock PLoS ONE, vol. 7, No.1.

\bibitem{slouf2014_sc}
Louf R., Barthelemy M., 2014, How congestion shapes cities : from mobility patterns to scaling
\newblock Scientific reports, Vol.4, No. 5561.

\bibitem{spumain2006}
Pumain, D., Paulus, F., Vacchiani-Marcuzzo, C., Lobo, J.,2006, An evolutionary theory for interpreting urban scaling laws
\newblock Cybergeo: European Journal of Geography, document 343.

\bibitem{sveregin1997}
Veregin, H., Tobler, W. R.,1997, Allometric relationships in the structure of street-level databases. 
\newblock Computers, environment and urban systems, 21(3), 277-290


\end{thebibliography}
\end{document}